\documentstyle[11pt]{article}

\begin{document}
\begin{center}
{\Large{\bf Roper Excitation in Alpha-Proton Scattering }}

\vspace{1.0truecm}

{S. Hirenzaki$^{a,}$
\footnote { JSPS research fellow }
, P. Fern\'andez de C\'ordoba$^b$, E. Oset$^a$ }

\vspace{1.0truecm}

{$^a$ Departamento de F\'{\i}sica Te\'orica and IFIC, Centro Mixto
\\Universidad de Valencia - CSIC,
46100 Burjassot (Valencia) Spain.\\}
{$^b$ Departamento de Matem\'atica Aplicada, Universidad Polit\'ecnica de
Valencia,
\\46022 Valencia, Spain.\\}

\vspace{2.0truecm}

{\bf Abstract\\}
\end{center}

We study the Roper excitation in the $(\alpha,\alpha')$ reaction.
We consider all
processes which may be relevant in the Roper excitation region, namely,
Roper excitation in the target, Roper excitation in the projectile, and
double $\Delta$ excitation processes.  The theoretical investigation
shows that the Roper excitation in the proton target mediated by
an isoscalar
exchange is the dominant mechanism in the process.
We determine an effective isoscalar interaction by means of which the
experimental cross section is well reproduced.  This should be useful
to make predictions in related reactions and is a first step to
construct eventually a microscopic $NN \rightarrow NN^*$ transition
potential, for which the present reaction does not offer enough
information.

\clearpage
{\bf 1 Introduction\\}

\vspace{0.1truecm}

We investigate theoretically the $(\alpha,\alpha')$ reaction on a proton target
at intermediate
energies in order to obtain new information on the reaction
mechanism and the properties of hadron resonances, especially the Roper
resonance. The fact that the $\alpha$ particle has isospin T=0 is particularly
useful, since, due to isospin conservation, it reduces the number of reaction
mechanisms which contribute to the reaction and allows an easier
interpretation of the results.

The experimental study of the $(\alpha,\alpha')$ reaction on the proton target
was done in ref. [1].  Two clear peaks were observed there; a large one, which
was associated in ref. [1] with $\Delta$ excitation in the projectile (DEP),
and a
small one, at higher excitation energies, which was attributed in ref. [1] to
the Roper excitation in the target.  This latter assumption requires the Roper
to be excited by the mediation of an isoscalar interaction
which stimulated the author of ref. [1] to interpret the Roper
resonance as a monopole excitation of the nucleon.

The idea of the DEP mechanism was suggested
theoretically in ref. [2] in connection with the $(^3 He,t)$ reaction
on nucleons and nuclei.  It was found there that this mechanism produced small
changes in the $(^3 He,t)$ reaction on proton targets with respect to the
dominant mechanism of $\Delta$ excitation in the target (DET), but the changes
were important in the reaction on neutron targets.  Thanks to this mechanism,
the excitation function of the $(^3 He,t)$ reaction on deuteron targets [3] was
well reproduced [4].  However, the clearest proof of the DEP mechanism was
found in the experiment of ref. [1] since, for reason of isospin conservation,
the DET mechanism is forbidden and all the strength for $\Delta$ excitation
comes from the DEP mechanism.  A theoretical study was done in ref. [5]
along these lines and the large peak corresponding to $\Delta$ excitation was
nicely reproduced.

Another interesting aspect of the work of ref. [5] is that it provides
an accurate tool to evaluate the "background" of the $(\alpha,\alpha')$
reaction which is necessary in order to
obtain the strength for the Roper excitation.
Given the fact that this background is much larger than the Roper signal, the
precise determination of the background is important in order to
asses the magnitude of the Roper excitation.  In ref. [1] some approximations
and assumptions were done to determine the shape of the $\Delta$ projectile
contribution, and the strength was fitted to reproduce the peak.  In ref.
[5] a more elaborate microscopic evaluation was done and both the shape
and magnitude of the cross section were determined.  As a consequence
there are some differences (not too large) in the $\Delta$ background
evaluated in refs. [1] and [5], and the strength of the Roper at its
peak is about $ 20 \% $ larger if the background of [5] is subtracted
instead of the one in [1].

In the present paper we study the different mechanisms that can lead to
the Roper excitation in the $(\alpha, \alpha')$ reaction on the proton.
However, instead of extracting the Roper signal by subtracting the
$\Delta$ background from the experimental cross section, we use the
theoretical model of ref. [5], which provides the $\Delta$ excitation,
and add to it the new mechanisms that excite the Roper.
This includes also the interference term between the target Roper and
the projectile $\Delta$ excitation, which are found to be the dominant
mechanisms.  With this global
model we obtain cross sections which are compared to the data in order to
extract new information on the Roper resonance. We find that the reaction
provides the strength
of an effective isoscalar exchange for the $NN \rightarrow NN^*$ transition.

In section 2 we calculate all processes which may be relevant in the
energy region of ref. [1], namely: Roper excitation in the target,
Roper excitation in the
projectile, and double $\Delta$ excitation process.  We compare the
calculated results with experimental data in section 3.   We summarize this
paper in section 4. \\

\vspace{0.3truecm}

{\bf 2 Model for the $(\alpha,\alpha')$ reaction\\}

\vspace{0.1truecm}

In this section we consider a theoretical model of the $(\alpha,\alpha')$
reaction on the proton target
in the $\Delta$ and Roper energy region. The reaction
mechanisms which we consider here are summarized in Fig. 1.
We include all processes which may be important in this energy region.
In Fig. 1 (a), we show the $\Delta$ excitation in the projectile.
Since the $\Delta$ can not be excited in the target [5],
this is the only process to excite the single $\Delta$ in the reaction.
We can find the detailed description of the
calculation and the results for this channel in ref. [5].
All the other channels are new and they are discussed below.

We consider the diagrams for the Roper resonance excitation depicted
in Figs. 1 (b-d).  In Fig. (b)
the Roper is excited in the target by the exchange of some isoscalar
objects between
the $\alpha$ and the proton.  Because of isospin conservation of the
$\alpha$, the isovector mesons ($\pi$ and $\rho$) do not contribute in this
process.  The cross section for this process is given by

\begin{equation}
   \frac{d^2{\sigma}}{dE_{\alpha'}d\Omega_{\alpha'}} =
   \frac{p_{\alpha'}}{(2{\pi})^3}
   \frac{2M_{\alpha}^2 M}{\lambda^{1/2}(s,M^2,M_{\alpha}^2)}
   \bar{\Sigma} \Sigma \vert T \vert ^2 \vert G^* (s^*) \vert ^2
\Gamma ^* (s^*) ,
\end{equation}

\noindent
where $\lambda(...)$ is the Kallen function and
$G^* (s)$ is the propagator of the Roper resonance defined as

\begin{equation}
  G^* (s) = \frac{1}{\sqrt{s} - M^* + \frac{i}{2} \Gamma ^* (s)} ,
\end{equation}

\noindent
where $M^*$ is the mass of the $N^*$, $M^{*} = 1440 MeV$
and $\Gamma ^* (s)$ is the energy dependent Roper width [6],

\begin{equation}
  \Gamma^{*} (s) = \Gamma^{*} (s=M^{*2})
  \frac{q_{cm}^3 (s)}{q_{cm} ^3 (M^{*2})} ,
\end{equation}

\noindent
with $\Gamma^{*} (s=M^{*2}) = 350 MeV$ and
$q_{cm} (s)$ the $\pi$ momentum in the center of mass frame of
$\pi N$ system with the energy $\sqrt{s}$.
Eq. (3) assumes for the $s$ dependence that the dominant decay channel
is $N^* \rightarrow \pi N$.
We will modify the width in the next section as described in the
Appendix in order to be more consistent with the experimental data.
In what follows, for simplicity, we construct a model assuming $\sigma$
exchange alone as responsible for the isoscalar part of the
$NN \rightarrow NN^*$ transition.  Further on we shall reinterpret the
meaning of this phenomenologically derived $"\sigma "$ exchange.
The spin sum and average of $\vert T \vert ^2$ can be written as

\begin{equation}
\bar{\Sigma} \Sigma \vert T \vert ^2 =
16 F^2 _{\alpha} g^2 _{\sigma NN^* } g^2 _{\sigma NN } \vert D_{\sigma} (q)
F^2_{\sigma} (q)\vert ^2 ,
\end{equation}

\noindent
where we are assuming couplings of the $\sigma$ to the $N$ and $N^*$ of the
type $g_{\sigma NN} \bar{\psi} \psi \phi$ and
     $g_{\sigma N N^*} \bar{\psi}_{N^*} \psi \phi + h.c.$.
In eq. (4) $D_{\sigma} (q)$ is the propagator of the $\sigma$-meson defined as

\begin{equation}
D_{\sigma} (q) = \frac{1}{ q^{02} - \vec{q}^{\: 2} - m^2 _{\sigma}} ,
\end{equation}

\noindent
with $m_{\sigma}$ = 550 MeV
, $F_{\sigma} (q)$ is the $\sigma$ form factor [7],

\begin{equation}
F_{\sigma} (q) = \frac{\Lambda ^2 _{\sigma} - m^2 _{\sigma}}
                      {\Lambda ^2 _{\sigma} - q^2          }
\end{equation}

\noindent
with $\Lambda _{\sigma}$ = 1700 MeV. In Eq. (4)
$F_{\alpha}$ is the $\alpha-\alpha'$ transition form factor which includes
the distortion effects and depends on the momentum transfer between
$\alpha$ and $\alpha'$.  The form factor is the same as that explained in ref.
[5] and accounts for the distortion of the nucleon wave plus the distortion
of a pion wave from the point of production of the pion.  It thus implicitly
assumes that the resonance will decay into the $\pi N$ system.
The pion distortion is slightly changed here.
We use the same eikonal form as in ref. [5] but take $Im \Pi = - p_{\pi}
 \sigma \rho$ with $\sigma$ the $\pi N$ experimental cross section and
$\rho$ the nuclear density.  This is appropriate at the higher energies
met in the present problem where the model of ref. [5] is not meant
to be applied.
The $\sigma NN$ coupling constant is taken from the
Bonn potential [7], $g^2 _{\sigma NN }/4 \pi = 5.69$,
and the $ \sigma NN^*$ coupling constant, $g _{\sigma NN^* }$,
is an unknown parameter which we shall determine from the experimental data.
We should however bare in mind that we are constructing an effective
isoscalar interaction and those couplings have not to be taken literally
as the meson baryon couplings of a microscopic model like in [7]. Yet
it is useful to take $g_{\sigma N N}$ as in the Bonn model since it already
provides the appropriate scale of the interaction strength.

In order to get eq. (1) we have replaced the energy conservation
$\delta$-function in terms of the Roper propagator and width as follows;

\begin{equation}
\delta (E_{\alpha} + E_N - E_{\alpha'} - E^*) \rightarrow
\frac{\Gamma ^* (s^*)}{2 \pi} \frac{E^*}{M^*} \vert G^* (s^*) \vert ^2 ,
\end{equation}

\noindent
so as to include all decay channels of the Roper resonance.

In the process shown in Fig. 1 (c), the Roper is excited in the projectile,
$\alpha$ particle, and decays into $\pi N$.  The Roper is excited by $\pi$
and $\rho$ exchange between the target and the projectile.  We include
both $\pi^+$ and $\pi^0$ for the final state.  We can write the cross section
as:

\begin{equation}
   \frac{d^2{\sigma}}{dE_{\alpha'}d\Omega_{\alpha'}} =
   \frac{p_{\alpha'}}{(2{\pi})^5}
   \frac{M_{\alpha}^2 M^2}{\lambda^{1/2}(s,M^2,M_{\alpha}^2)}
   \int d^3 p_{\pi}
   \frac{1}{E_{N'} \omega_{\pi}}
   \bar{\Sigma} \Sigma \vert T \vert ^2
   \delta (E_{\alpha} + E_N - E_{\alpha'} - E_{N'} - \omega_{\pi})  .
\end{equation}

\noindent
The spin sum and average of $\vert T \vert ^2$ for this process is ;

\begin{displaymath}
\bar{\Sigma} \Sigma \vert T \vert ^2 =
48 F_{\alpha} ^2 \left(\frac{f}{\mu} \right)^2 \left(\frac{f'}{\mu} \right)^4
\vert G^* (s^*) \vert ^2
\end{displaymath}

\begin{equation}
\times
[(V_l ^{\prime 2} (q) - V_t^{\prime 2} (q))(\vec{p}_{\pi CM} \cdot \hat{q} )^2
+ V _t^{\prime 2} (q) \vec{p}_{\pi CM} ^{\: 2}]
\left( \frac{-q^2}{\vec{q}^{\: 2}} \right)  ,
\end{equation}

\noindent
where $q = p_N - p_{N'}$, $\vec{p}_{\pi CM}$ is the pion momentum in
the Roper rest frame and $f^2/4\pi = 0.08$, $f' = 0.472$ [6].
The factor $(- q^2 / \vec{q} ^{\: 2} ) $ arises from the relativistic invariant
$\pi NN$ vertex [5].
$V_l ^{\prime}$, $V_t ^{\prime}$ stand for the longitudinal and
transverse part of the $NN \rightarrow NN^*$ interaction.
We have taken,

\begin{equation}
V_l ^{\prime} (q) = \left( \frac{\vec{q} ^{\: 2} }
              {q^{02} - \vec{q}^{\: 2} - {\mu} ^2}
                           F^2 _{\pi} (q) + g' \right)
\end{equation}

\begin{equation}
V_t ^{\prime} (q) = \left( \frac{\vec{q} ^{\: 2} }
             {q^{02} - \vec{q}^{\: 2} - m_{\rho} ^2}
                           F^2 _{\rho} (q) C_{\rho}+ g' \right)   ,
\end{equation}

\noindent
where $F_{\pi} (q)$ and $F_{\rho} (q)$ are the pion and $\rho$-meson form
factor in the form of eq. (6) with $\Lambda_{\pi} = 1300 MeV$ and
$\Lambda_{\rho} = 1400 MeV$, $C_{\rho} = 3.96$ [7], and $g'$, the
Landau-Migdal parameter, is taken to be 0.60.  The momentum $q$ in
eqs. (10), (11) are taken in the Roper rest frame [5].  The invariant
mass $\sqrt{s^*}$ of the Roper is approximated to be

\begin{equation}
s^* = (q^{0} + M )^2 - \left( \frac{\vec{q}+\vec{p}_{\pi}}{2} \right)^2
\end{equation}

\noindent
using the momentum variables in the $\alpha$ rest frame [5].
In this approximation the momentum transfer is
shared equally by the initial and final nucleon in the $\alpha$.

Now we consider the process of Fig. 1 (d), the projectile Roper
excitation which decays into the nucleon and the two pions in the $T=0$,
$S$-wave channel, which carries a certain fraction of the Roper width [8].
We have again only the isoscalar exchange contribution because of isospin
conservation, which is accounted for by means of the effective $\sigma$
exchange used for diagram (b).
The cross
section can be expressed as,

\begin{displaymath}
   \frac{d^2{\sigma}}{dE_{\alpha'}d\Omega_{\alpha'}} =
   \frac{p_{\alpha'}}{2 (2{\pi})^8}
   \frac{M_{\alpha}^2 M^2}{\lambda^{1/2}(s,M^2,M_{\alpha}^2)}
   \int d^3 p_{\pi_{2}}
   \frac{1}{\omega_{\pi_{2}}}
   \int d^3 p_{\pi_{1}}
   \frac{1}{E_{N'} \omega_{\pi_{1}}}
\end{displaymath}

\begin{equation}
   \times
   \bar{\Sigma} \Sigma \vert T \vert ^2
   \delta (E_{\alpha} + E_N - E_{\alpha'} - E_{N'}
    - \omega_{\pi_{1}} - \omega_{\pi_{2}})  .
\end{equation}

\noindent
The spin sum and average of $\vert T \vert ^2$ is now,

\begin{equation}
\bar{\Sigma} \Sigma \vert T \vert ^2 =
\frac{3}{2}
64 F^2 _{\alpha} C^2 g_{\sigma NN } ^2 g_{ \sigma NN^* } ^2
\vert G^* (s^*) \vert ^2
\vert D_{\sigma} (q) F^2 _{\sigma} (q) \vert ^2 ,
\end{equation}

\noindent
where $C$ is the coupling constant of the $N^* \rightarrow N + 2\pi$ decay and
$C = -2.66 \mu ^{-1}$ [6]. The variable $s^*$ is obtained in a similar way
as in eq. (12),

\begin{equation}
s^* = (q^{0} + M )^2 -
\left( \frac{\vec{q}+\vec{p}_{\pi_1}+\vec{p}_{\pi_2}}{2} \right)^2
\end{equation}

\noindent
with the momenta in the $\alpha$ rest frame.

We omit details of the effective Lagrangians and couplings used for the
different vertices. All of them are compiled in appendices A and B of ref. [6]
and we follow them strictly.  The factor $\frac{3}{2}$ in front of eq. (14)
is an isospin factor which sums the contribution of the $\pi^+ \pi^-$ decay
channel and the $\pi^0 \pi^0$ decay channel (which has the factor $\frac{1}{2}$
of symmetry).

In addition to this decay channel we could add the $N^* \rightarrow \pi \Delta$
channel which carries a fraction of $20-30 \% $ of the $N^*$ decay width.
However, as we shall see, the projectile Roper excitation mechanism with the
dominant $N^*$ decay channel, $N^* \rightarrow \pi N$ (Fig. 1(c)), which we
have studied before, gives a negligible contribution, basically because of
the small $\pi N N^*$ coupling.  Since in this case one has again the
exchange of $\pi$ and $\rho$ mesons as in Fig. 1(c), and the fraction of the
$N^* \rightarrow \pi \Delta$ decay is smaller than that of $N^* \rightarrow
\pi N$, this mechanism should give even a smaller contribution and
we do not evaluate it here.

Finally we consider the double $\Delta$ excitation process as shown in
Fig. 1 (e).  We have $\pi$ and $\rho$ meson exchange in this process and
we have two $\Delta$ resonances, one is in the target and the other one in the
projectile.  The cross section is,

\begin{equation}
   \frac{d^2{\sigma}}{dE_{\alpha'}d\Omega_{\alpha'}} =
   \frac{p_{\alpha'}}{ (2{\pi})^6}
   \frac{M_{\alpha}^2 M}{\lambda^{1/2}(s,M^2,M_{\alpha}^2)}
   \int \frac{d^3 p_{\pi}}{\omega_{\pi}}
   \bar{\Sigma} \Sigma \vert T \vert ^2
   \vert G_{\Delta_{T}} (s_{\Delta_{T}})\vert ^2
   \Gamma_{\Delta_{T}} (s_{\Delta_{T}})
\end{equation}

\noindent
where the propagator and the width of the $\Delta$ are defined as,

\begin{equation}
G_{\Delta} (s) = \frac {1}{\sqrt{s} - M_{\Delta} +
\frac{i}{2} \Gamma_{\Delta} (s)}  ,
\end{equation}

\noindent
and

\begin{equation}
\Gamma_{\Delta} (s) =
\frac{2}{3} \frac{1}{4 \pi} \left( \frac{f^*}{\mu} \right)^2
\frac{M}{\sqrt{s}} q^3 _{cm}
\end{equation}

\noindent
with $M_{\Delta}=1232 MeV$, $f^{*2}/4 \pi = 0.36$
and $q_{cm}$ the $\pi$ momentum for $\Delta$ decay at rest with
mass $\sqrt{s}$ in the $\pi N$ system. The index $\Delta_{T}$
indicates the $\Delta$ resonance in the target.  Here we replaced the energy
conservation $\delta$-function in terms of the $\Delta$ propagator and
the width in the target in the same way as eq. (7).  The sum and average
over spin of $\vert T \vert ^2$ is given as,

\begin{equation}
\bar{\Sigma} \Sigma \vert T \vert ^2 =
\left( \frac{16}{9} \right) ^2 \frac{4}{3} F^2 _{\alpha}
\left( \frac{f^*}{\mu} \right) ^6
\vert G_{\Delta_{P}} (s_{\Delta_{P}}) \vert ^2
[(V_l ^{\prime 2} (q) - V_t^{\prime 2} (q))(\vec{p}_{\pi CM} \cdot \hat{q} )^2
+ V _t^{\prime 2} (q) \vec{p}_{\pi CM} ^{\: 2}]  ,
\end{equation}

\noindent
where $V'_l, V'_t$ are defined in eqs. (10) and (11).  The index $\Delta_{P}$
indicates the $\Delta$ resonance in the projectile. The magnitude of
$s_{\Delta_{P}}$ is defined as eq. (12). Eq. (19) already accounts for the
possibility of $\pi^0 , \pi^+ , \pi^-$ decay of the projectile $\Delta$
and all isospin channels of the target $\Delta$.

As we shall see later on, the diagrams of Figs. 1 (c), (d), (e) are negligible
and the two important mechanisms are given by the diagrams of Figs. 1 (a),
(b).  When we compare our calculated results with the data [1], we include the
interference of the two processes.
Obviously we must select only the $N^* \rightarrow \pi N$ decay channel
in Fig. 1 (b) and evaluate the amplitude for this process explicitly
in order to have the same final state as in Fig. 1 (a) and thus have
some interference.  The interference contribution is given
by eq. (8) replacing $\bar{\Sigma} \Sigma \vert T \vert ^2$ by

\begin{displaymath}
\bar{\Sigma} \Sigma ( T^* _{N^*} T_{\Delta} + T^* _{\Delta} T_{N^*} )
\end{displaymath}

\begin{displaymath}
= 2 Re \left[ \frac{64}{3} F^2 _{\alpha}
 \left( g_{\sigma NN} F_{\sigma} D_{\sigma} g_{\sigma NN^*} F_{\sigma}
G^* \frac{f'}{\mu} \right)^*
 \right.
\end{displaymath}

\begin{equation}
 \times
 \left.
 \left( \frac{f^*}{\mu} G_{\Delta} \frac{f^*}{\mu}
         [(V_l ^{\prime} - V_t^{\prime})
     (\vec{p}_{\pi \Delta} \cdot \hat{q} )(\vec{p}_{\pi *} \cdot \hat{q} )
         + V_t^{\prime} (\vec{p}_{\pi \Delta} \cdot \vec{p}_{\pi *})]
        \frac{f}{\mu} \sqrt{\frac{-q^2}{\vec{q}^{\: 2}}} \right) \right]
\end{equation}

\noindent
where $T_{N^*}$ is the $T$ matrix of the target Roper process followed by
$\pi N$ decay, $T_{\Delta}$ is that of the projectile $\Delta$ process,
and $\vec{p} _{\pi \Delta}$ is the pion momentum in the $\Delta$ rest frame
and $\vec{p} _{\pi *}$ is in the $N^*$ rest frame.  This last expression sums
the contribution from the production of a $\pi^0$ and a $\pi^+$.

We should note that the interference between the $T=1/2$ and $T=3/2$
excitations (with the simultaneous different spin excitation)
has appeared because they occur on different nucleons, one in the target
and the other one in the projectile.  Should these excitations had occured
both on the target nucleon there would have been no interference.
In our case the $\Delta$ excitation in the target is forbidden but
it would have appeared if we had a $^3 He$ projectile instead of $^4 He$,
and there would be no interference between $\Delta$ excitation and
Roper excitation on the target.
\\

\vspace{0.3truecm}

{\bf 3 Numerical Results\\}

\vspace{0.1truecm}

We should mention first the
gross features of the data.  As can be seen in ref. [1], the observed cross
section has a peak around $\omega = 550 MeV$ after subtracting the
contribution of the projectile $\Delta$ excitation ( Fig. 1 (a))
of ref. [1], which indicates the Roper excitation [1].
The data of the energy integrated cross section of this $N^*$
peak are also available at several angles [1]. The data of ref. [1] has been
reanalysed with a more precise background subtraction [9].  With these
corrections the height at the $\Delta$ peak is about $15 \% $ lower than
in ref. [1]. In Fig. 2 we show the new spectrum [9] with the appropriate
normalization deduced from the scales in the energy integrated cross section
of ref. [1] and the correction in ref. [9].  By subtracting the $\Delta$
background evaluated in ref. [5] we can see that the strength of the
Roper excitation at its peak is of the order of 0.25 [mb/sr/MeV].

We evaluate the cross section with the mechanisms discussed in the former
section and show the results in Fig. 3.  As we said, in the diagrams Fig. 1(c)
and Fig. 1(e) all the couplings are known.  Hence, we can calculate the cross
section from these diagrams, which we show in the figure.  As we can see there,
their strength is very small and by no means can they account for the
strength in the Roper region.  This leaves diagrams Fig. 1(b) and 1(d) to do
the job.  The cross sections for these two processes are both proportional
to $g^2 _{\sigma N N^*}$.  Even without knowing anything about this coupling,
we can determine the ratio of the cross sections for these two mechanisms.
We found that the target Roper process is much more important than the
projectile Roper process followed by $\pi \pi N$ decay by about a factor 100.
The cross section of the projectile process is suppressed because of the final
state phase space which involves two pions.

Hence, diagram Fig. 1(b) for Roper excitation in the target stands as the
only likely mechanism to explain the data.  Thus we fix for the moment the
strength of $g _{\sigma N N^*}$, the only unknown in the theory, in order to
reproduce a strength of the peak of about 0.25 [mb/sr/MeV].  The value of the
coupling constant that we get is $g^2 _{\sigma N N^*} / 4 \pi = 1.79$.  With
this coupling we can now evaluate the diagram Fig. 1(d) and we find, as
shown in the figure, a very small contribution.

We can explain the reasons why those terms are so small
here.  The cross section of the projectile Roper excitation can be
compared with that of the projectile $\Delta$ excitation (Fig. 1(a))
in ref. [5] directly. They have the same phase space and the same $T$ matrix
except for some factors.
We found that the cross section is so small simply because of the small
coupling constants. The cross section of the projectile Roper
excitation can be evaluated from that of projectile $\Delta$ excitation
using a ratio of the coupling constants, $(f'/f^*)^4 = 2.4 \times 10^{-3}$.

For the double
$\Delta$ process the reasons are the following: first, the peak position
of the target $\Delta$ excitation is different from that of the projectile
$\Delta$ excitation because of the kinematics [2,4].
Hence, the cross section is
the result of a small overlap of two different resonance peaks.
Second, the resonant strength associated to $\Delta$ excitation in the
projectile,
which peaks at small excitation energies, is now considerably reduced because
the phase space available is very restricted when one forces another
$\Delta$ to be excited simultaneously in the target.
To confirm our results we try to evaluate the result of the double $\Delta$
process using the available ones, from that of the projectile $\Delta$ process.
The $T$ matrix is the same in both processes except for some factors.
The phase space is now different due to the different final states.
To simulate the double $\Delta$ process we increase the
final nucleon mass of the projectile $\Delta$ process.  We found that the
projectile $\Delta$ process with 940+250 [MeV] final nucleon mass has a peak
at the same position of that of the double $\Delta$ process, and its height is
around 1/100 of the original projectile $\Delta$ process because of the
phase space differences.
In addition the peak height of the double $\Delta$ process must be even
lower than this peak because of the $\Delta$ width in the target. Hence,
we can reconfirm qualitatively the small
contribution of the double $\Delta$ process.

All there things considered, the Roper excitation in the target of Fig. 1 (b)
is the only mechanism which is left to explain the data.
All other processes
(Fig.1 (c-e)) provide typically two orders of magnitude smaller cross
section than the experimental data. As we can see in the figure, we need only
the target Roper excitation and we neglect all the other processes hereafter,
except for the projectile $\Delta$ excitation which is large and
has already been evaluated [5].

We show the target Roper contribution together with the projectile $\Delta$
contribution [5] and their interference
in Fig. 2 and compare them to the data.
Here we take the $g ^2 _{ \sigma N N^{*} } / 4 \pi = 2.35 $.  We found that
the Roper excitation produces a wide peak around $\omega = 520 MeV$.
The interference has a
negative contribution to the cross section and peaks around
$\omega = 350 MeV$.
The calculated cross section provides a fair account of the cross section
but the dip region between $N^*$ and $\Delta$ excitation is poorly reproduced.
We have chosen a particular sign for $g_{\sigma N N^*}$, the same as
$g_{\sigma N N }$, which leads to destructive interference.  If the opposite
sign is chosen, the constructive interference leads to a cross section in
large disagreement with the data.

In order to obtain a better agreement with the data we change
the expression of the width of the Roper resonance in eq. (3).
Experiments tell us that the Roper resonance decays not only into $ \pi +  N$
$ (65\%)$
but also into $ \pi + \pi +  N$ $ (35\%) $ [8]. We describe in the Appendix
how we take into account the $ 2 \pi + N$ decay.  The Roper width
$\Gamma^{*}$ in eqs.
(1) and (2) is replaced by this new form and the distortion effects of
final $ 2\pi$ are also considered in $F_{\alpha}$.
Then we take the freedom to change the Roper mass and width in the
range of their uncertainties [8] and try to obtain a best fit to the data
by changing $M^*, \Gamma^* (s=M^{*2})$, and $g_{\sigma NN^*}$.
The calculated results depend generally on these parameters
in the following way: the peak moves to a lower $\omega$ value for larger width
and smaller mass, the peak is higher for smaller width and larger $g_{\sigma
NN^*}$, the peak is steeper for smaller width, and the interference is
relatively more
important for smaller $g_{\sigma NN^*}$.  The result for our best fit
is shown in Fig. 4, where we see that the data are well reproduced.
The best fit parameters have been : $M^* = 1430 MeV, \Gamma^* (s=M^{*2}) =
300 MeV$, and $g^2 _{\sigma N N^*} /4 \pi =1.33$.

We show the calculated angular distribution of the Roper excitation in Fig. 5.
The interference contribution is not included in this distribution.
The data are from ref. [1] and they should be corrected by the new background
subtraction [9].  We should also notice that the fact that the interference
term between the projectile $\Delta$ and target Roper mechanism is not small
does not allow a clean experimental separation of these mechanisms.  With
this caveat, the comparison of our results with the experimental data should
only be taken as qualitative.  The main point we want to stress here is that
the monotonous fall down of the cross section is reproduced and, in our
theoretical analysis, it is mostly a consequence of the $(\alpha,\alpha')$
transition form factor and not a property tied to the Roper itself.  We
found that our results reproduce the trend of the data well.

Finally we want to comment on the $\pi N$ scattering amplitude of $P_{11}$
channel.  In this channel the observed amplitude [11, 12] has a different
form than the standard Breit-Wigner form
of the Roper resonance due to the coupling to the nucleon.  In the energy
region which we consider in this paper, the differences are as follows;
first the real part of the observed amplitude has the opposite sign to the
Breit-Wigner form at $\sqrt{s} \leq 1.2 GeV$ and second the shape of
the real part of the observed scattering amplitude is steeper than
the Breit-Wigner form at $1.2 \leq \sqrt{s} \leq 1.3 GeV$ because of the
off-shell nucleon effect . In order to see the effect of these
differences we calculated the $\alpha$ spectrum with a modified Roper
propagator which has a steeper real part at $1.2 \leq \sqrt{s} \leq 1.3 GeV$
according to the data of the scattering amplitude.  We have checked that
including these modifications in the "Roper" excitation changes only a bit
the results of Fig. 4 in the region of the dip, reducing moderately the
cross section there.  Theoretically the inclusion of the nucleon pole term
in addition to the Roper pole would help producing the shape in the $P_{11}$
channel.

Now we would like to comment on the meaning of the $"\sigma "$ exchange
interaction obtained.  In a more microscopic description of the
$NN \rightarrow NN^*$ transition along the lines of the boson exchange
model,  in the isoscalar channel which we have investigated we would
also have a contribution from $\omega$ exchange and from uncorrelated
$2\pi$ exchange.  It is easy to see that assuming a similar scaling here
for $\omega$ exchange and the uncorrelated $2\pi$ exchange, with respect to
$\sigma$ exchange, as one has in the $NN$ potential [7], the effect of
$\omega$ and uncorrelated $2\pi$ exchange are very important and one
finds large cancellation between $\sigma$ and $\omega$ exchange.  In addition
one should use this as input for a transition potential and initial and
final state interactions of the $NN$ or $NN^*$ systems (correlations) should
also be taken into account.  For all these reasons the $"\sigma "$ exchange
potential which we have obtained should not be interpreted as a $\sigma$
exchange for the $NN \rightarrow NN^*$ transition along the lines of a
one boson exchange model.  It is simply an effective interaction which
accounts for all the ingredients in the $T=0$ exchange channel, ($\sigma$,
$\omega$ and correlations).  One may wonder why using there the explicit
$\sigma$ mass in the exchange.  There is certainly no justification for it,
except that a posteriori one finds that the mass of the object exchanged is
irrelevant in the description of the cross section and it can be equally
reproduced using any other mass.  Hence the $"\sigma"$ exchange obtained
stands only as a useful and intuitive parameterization of the effective
interaction in the $T=0$ channel.  With this easy interaction one can
make predictions for analogous reactions using other nuclei, one can
evaluate cross sections at other energies of the beam, etc.

Obviously, although the limited information of the present reaction does not
allow one to extract enough information to construct a one boson exchange
model for the $NN \rightarrow NN^*$ transition, the job done here,
separating the $\Delta$ projectile excitation from the Roper excitation,
provides some partial, but useful information, on the $NN \rightarrow
NN^*$ transition to be used in the future in attempts to construct
a microscopical model for this interaction.  Some steps in this
direction, by looking at the role of uncorrelated $2\pi$ exchange, have been
given in ref. [13].

\vspace{0.3truecm}

{\bf 4 Summary\\}

\vspace{0.1truecm}

We have studied the Roper excitation in the $(\alpha,\alpha')$ reaction on the
proton target.  All processes which may be relevant in this energy region
were investigated.  We found that the experimental $\alpha'$ spectrum can be
reproduced by two processes, the projectile $\Delta$ excitation and
the target Roper process. The target Roper process is mediated by an
isoscalar
exchange between the $\alpha$ and the proton and we have determined from the
experiment the effective isoscalar $NN \rightarrow NN^*$ transition
$t$ matrix.

We could find a good reproduction of the data with values of $M^*$ and
$\Gamma ^* $ close to the average values quoted in the particle
data table [8].  We found a good agreement with the data with $M^* =
1430 MeV$, $\Gamma^* (s=m^{* 2}) = 300 MeV$ and a certain choice of
the parameters of the effective interaction.

The experimental dependence of the cross section on the $\alpha'$
angle was qualitatively reproduced and found to be tied to the $\alpha$
form factor, not to the properties of the Roper.

The present work also lays the ground for extension of studies of $N^*$
excitation in nuclei in order to study the modification of the $N^*$
properties in a nuclear medium.  The excitation of the $N^*$ with the
$(\alpha,\alpha ')$ reaction, because of the large strength and clean
signature, would be probably one of the ideal tools for such studies.

\noindent
{\bf Acknowledgements}

We would like to thank useful discussions with E. Hern\'andez,
M. J. Vicente-Vacas and H. P. Morsch.
We would also like to thank the latter for providing us with the results of the
reanalysis of the experiment.  One of us S. Hirenzaki
would like to acknowledge the hospitality of the University of Valencia
where this work was done.  This work is supported partially by
CICYT contract number AEN-93-1205. \\

\clearpage

{\appendix
{\bf \quad Appendix Decay Width of the Roper resonance}

\vspace{0.1truecm}
In this appendix we will explain our model of the widths of the Roper
resonance.  We include the $N^* \rightarrow  \pi + N$ and
$N^* \rightarrow  \pi + \pi + N$ decay channels.
Writing the decay width of each channel by $\Gamma^* _{ \pi}$ and
$\Gamma^* _{ \pi \pi}$, we define the total decay width as,

\begin{equation}
 \Gamma^{*} (s) = \Gamma ^*_{ \pi} (s) + \Gamma ^*_{ \pi \pi} (s)  .
\end{equation}

The width of the $\pi + N$ decay channel has the same form as that of
ref. [6],

\begin{equation}
  \Gamma^* _{ \pi} (s) = \Gamma^* _{ \pi} (s=M^{*2})
  \frac{q_{cm}^3 (s)}{q_{cm} ^3 (M^{*2})} ,
\end{equation}

\noindent
where $\Gamma^* _{ \pi} (s=M^{*2}) = P_{\pi} \Gamma^{*} (s=M^{*2}) $,
with $\Gamma^{*} (s=M^{*2})$ the experimental Roper width
and $ P_{\pi}$ the $\pi N$ decay branching ratio.
The magnitude $q_{cm} (s)$ is the $\pi$ momentum in the center
of mass frame of the
$\pi N$ system with energy $\sqrt{s}$.

For the width of the $\pi +  \pi + N$ decay channel, we assume the
$ N^{*} \rightarrow \pi + \Delta $ as an intermediate state in this paper and
express the width as follows,

\begin{equation}
  \Gamma^* _{ \pi \pi} (s) = \int
                   \frac{d^3 p_{\pi}}{(2\pi)^3}
                   \frac{d^3 p_{\Delta}}{(2\pi)^3}
                   \frac{M_{\Delta}}{E_{\Delta}}
                   \frac{1}{2 \omega_{\pi}}
                   \bar{\Sigma} \Sigma \vert T \vert ^2
                   (2 \pi)^4 \delta ^4 (p^{*} - p_{\pi} - p_{\Delta}) ,
\end{equation}

\noindent
where $p^{* \mu}  $ is the four momenta of the Roper resonance and is
$ ( \sqrt{s} , \vec{0} )$ in the Roper rest frame.
The $\pi \Delta N^{*}$ coupling
is taken to be of the same form that of $\pi N \Delta $ with the coupling
strength $f_{\pi \Delta N^{*}} $ [6].  After replacing the energy conservation
$\delta$-function into the $\Delta$ propagator and width as in eq. (7) in the
text, we find the $\Gamma^* _{ \pi \pi}$ is described as,

\begin{equation}
\Gamma^* _{ \pi \pi} (s) = \frac{1}{3 \pi ^2}
   \left( \frac{f_{\pi \Delta N^{*}}}{\mu} \right)^2
   \int d p_{\pi}
   \frac{p_{\pi}^4}{\omega_{\pi}}
   \vert G_{\Delta} (s_{\Delta}) \vert ^2
   \Gamma_{\Delta} (s_{\Delta})  ,
\end{equation}

\noindent
which has included all the isospin channels,
where $G_{\Delta}$ and $\Gamma_{\Delta}$ are defined in eqs. (17) and (18),
respectively.  The coupling constant, $f_{\pi \Delta N^{*}}$, is determined
by the normalization condition,
$\Gamma^* _{ \pi \pi} (s=M^{*2}) = P_{\pi \pi} \Gamma^{*} (s=M^{*2}) $
with $\Gamma^{*} (s=M^{*2})  $ the experimental Roper width and $ P_{\pi \pi}$
 the $\pi \pi N$
decay branching ratio.  We obtain $f_{\pi \Delta N^*} =
 2.47 $ for  $M^*=1440 MeV, \Gamma ^* (s=M^{*2}) = 350 MeV
$ and $P_{\pi \pi} =0.35$ [8].
}

\clearpage
{\bf References\\}

\vspace{0.3truecm}

\noindent
[1] H. P. Morsch $et$  $al$., Phys. Rev. Lett. {\bf 69} (1992) 1336.

\noindent
[2] E. Oset, E. Shiino, and H. Toki, Phys. Lett. {\bf B224} (1989) 249.

\noindent
[3] C. Gaarde, Nucl. Phys. {\bf A478} (1988) 475c.

\noindent
[4] P. Fern\'andez de C\'ordoba and E. Oset, Nucl. Phys. {\bf A544} (1992) 793.

\noindent
[5] P. Fern\'andez de C\'ordoba, E. Oset, M. J. Vicente-Vacas, Yu. Ratis,
J. Nieves, B. L\'opez-Alvaredo, and F. Gareev, Nucl. Phys. {\bf A586} (1995)
586.

\noindent
[6] J. A. G\'omez Tejedor and E. Oset, Nucl. Phys. {\bf A571} (1994) 667.

\noindent
[7] R. Machleidt, K. Holinde, and Ch. Elster, Phys. Rep. {\bf 149} (1987) 1.

\noindent
[8] Review of Particle Properties, Phys. Rev. {\bf D50} (1994) 1173.

\noindent
[9] H. P. Morsch, private communication.

\noindent
[10] K. Holinde, private communication.

\noindent
[11] R. A. Arndt, J. M. Ford, L. D. Roper, Phys. Rev. {\bf D32} (1985) 1085.

\noindent
[12] R. E. Cutkosky and S. Wang, Phys. Rev. {\bf D42} (1990) 235.

\noindent
[13] B. Desplanques, Z. Phys. {\bf A330} (1988) 331.

\clearpage

{\bf Figure Caption\\}

\vspace{0.3truecm}

\noindent
{\bf Fig. 1} \quad Diagrams for the $(\alpha,\alpha')$ reaction which
we consider in this paper.  They are (a) $\Delta$ excitation in
the projectile calculated in ref. [5], (b) Roper excitation in the
target, (c) Roper excitation in the projectile with decay into $\pi N$,
(d) Roper excitation in the projectile with decay into $\pi\pi N$, and
(e) double $\Delta$ excitation.  The $\sigma$ exchange must be
interpreted as an effective interaction in the $T=0$ exchange
channel (see text).

\noindent
{\bf Fig. 2} \quad Calculated cross sections of the target Roper
process and the projectile $\Delta$ process [5] at E$_{\alpha}$= 4.2 GeV
and $\theta$ = 0.8 deg.  The variable $\omega$
is the energy transfer defined as
$\omega = E_{\alpha} - E_{\alpha '} $.  The thick line
indicates the sum of all contributions. Experimental data are taken from
ref. [11]. Here we used $g ^2_{ \sigma N N^{*}}/ 4 \pi = 2.35$.

\noindent
{\bf Fig. 3} \quad Calculated cross sections $d\sigma /d\Omega dE$ for
$(\alpha ,\alpha ')$ on the proton at E$_{\alpha}$= 4.2 GeV
and $\theta$ = 0.8 deg.  The variable $\omega$
is the energy transfer defined as
$\omega = E_{\alpha} - E_{\alpha '} $.
Each line indicates the contribution from
the process shown in Fig. 1.  Here we used $g ^2_{ \sigma N N^{*}}
/ 4 \pi = 1.79$.

\noindent
{\bf Fig. 4} \quad Same as in Fig. 2.
Here we used $g ^2_{ \sigma N N^{*}}/ 4 \pi = 1.33, M^*= 1430MeV,
\Gamma^* (s=M^{*2})= 300MeV$ and the Roper width discussed in the appendix.

\noindent
{\bf Fig. 5} \quad Calculated differential cross sections, $d\sigma /d\Omega$,
of the target Roper process as a function of the scattering angle in the
laboratory frame.  The parameters are the same as in Fig. 4.
The experimental data are taken from ref. [1]. See warnings in the text
about the interpretation of the results.
\end{document}